\documentclass[aps,prd,nofootinbib,showkeys,floatfix,notitlepage]{revtex4-1}
\usepackage{bbold}
\usepackage{amsmath}
\usepackage{amssymb}
\usepackage{graphicx}
\usepackage{xspace}
\usepackage{bbm} 
\usepackage{color}
\usepackage[utf8]{inputenc}
\usepackage{cancel}
\usepackage{slashed}
\usepackage{soul}
\usepackage{units}
\usepackage{enumitem}
\usepackage{todonotes}
\usepackage{soul}

\usepackage{tikz} 
\usetikzlibrary{chains,shapes,fit,calc}
\usetikzlibrary{positioning,arrows}

\usetikzlibrary{patterns,decorations.pathreplacing}
\usetikzlibrary{decorations.pathmorphing}
\usetikzlibrary{decorations.markings}
\usetikzlibrary{arrows,shapes}
\usetikzlibrary{shapes.misc}
\tikzset{
    gluon/.style={decorate, draw=black,
        decoration={coil,amplitude=4pt, segment length=4pt,aspect=0.7}} 
}
\tikzset{
    photon/.style={decorate, decoration={snake}},
}

\newcommand\higgsb{{\tt HiggsBounds}\xspace}
\newcommand\higgss{{\tt HiggsSignals}\xspace}
\newcommand\ScannerS{{\tt ScannerS}\xspace}
\newcommand\BSMPT{{\tt BSMPT}\xspace}

\newcommand{\Lag}{\mathcal{L}}
\newcommand{\LagBSM}{\Lag^\text{BSM}_\text{fermion}}

\newcommand{\msusy}{M_{\tiny SUSY}}
\newcommand{\vev}{\textit{v}}

\newcommand{\xithdm}{\xi^{\text{2HDM}}_c}

\newcommand{\VCW}{V_\text{CW}}
\newcommand{\VD}{V_\text{\tiny Debye}}
\newcommand{\VT}{V_{\text{T}}}

\newcommand{\cb}{c_\beta}
\renewcommand{\sb}{s_\beta}

\newcommand{\ms}[1]{m_{#1}^2}

\newcommand{\lam}[1]{\lambda_{#1}}
\newcommand{\p}[1]{\Phi_{#1}}

\def\tW{\tilde{W}}
\def\tB{\tilde{B}}
\def\tHu{{\tilde{H}_u}}
\def\tHd{{\tilde{H}_d}}
\def\Hu{{H_u}}
\def\Hd{{H_d}}

\newcommand{\s}{\newline \vspace*{-2.5mm}}

\usepackage[pdftex,colorlinks,linkcolor=cyan,citecolor=blue,urlcolor=cyan]{hyperref} 
\hypersetup{linktocpage}
\usepackage[capitalise, english]{cleveref}
\crefname{chapter}{Chapter}{Chapter}
\crefname{section}{Sec.}{Secs.}
\crefname{table}{Tab.}{Tabs.}
\crefname{figure}{Fig.}{Figs.}
\crefname{equation}{Eq.}{Eqs.}
\crefname{appendix}{Appendix\ }{Appendix\ }
\Crefname{section}{Section}{Sections}
\Crefname{table}{Tale}{Tables}
\Crefname{figure}{Figure}{Figures}
\Crefname{equation}{Equation}{Equations}

\graphicspath{{figs/}}
\makeatletter
\def\input@path{{chapters/}}
\makeatother

\begin{document}
\vspace{1cm}

\title{\LARGE Electroweak Phase Transitions with BSM Fermions}

\hfill \parbox{5cm}{\vspace{ -1cm } \flushright KA-TP-16-2021}

\newcommand{\AddrKAITP}{
Institute for Theoretical Physics (ITP), 
Karlsruhe Institute of Technology, \\
Wolfgang-Gaede-Stra{\ss}e 1, D-76131 Karlsruhe, Germany}

\newcommand{\AddrKAIKP}{
Institute for Nuclear Physics (IKP), Karlsruhe Institute of Technology,\\
Hermann-von-Helmholtz-Platz 1, D-76344 Eggenstein-Leopoldshafen, Germany}

\author{Martin Gabelmann} \email{martin.gabelmann@kit.edu} 
\affiliation{\AddrKAITP}

\author{M. Margarete M\"uhlleitner} \email{milada.muehlleitner@kit.edu} 
\affiliation{\AddrKAITP}

\author{Jonas M\"uller}\email{jonas.mueller@kit.edu}
\affiliation{\AddrKAITP}

\begin{abstract}
   We study the impact of additional beyond-the-Standard Model (BSM)
    fermions, charged under the Standard Model (SM) 
    $SU(2)_L \otimes U(1)_Y$ gauge group, on the electroweak
    phase transition (EWPT) in a 2-Higgs-Doublet-Model (2HDM) of
    type II.
    We find that the strength of the EWPT can be enhanced by about 40\%
    compared to the default 2HDM. 
    Therefore, additional light fermions are a useful tool to weaken the
    tension between increasing mass constraints on BSM scalars and the
    requirement of additional light scalar degrees of freedom to
    accommodate a strong first order EWPT.
    The findings are of particular interest for a variety of
    (non-minimal) split supersymmetry scenarios which necessarily introduce
    additional light fermion degrees of freedom.
\end{abstract}

\maketitle

\section{Introduction}
\label{sec:intro}
The discovery of a Higgs boson by the LHC experiments ATLAS
\cite{Aad:2012tfa} and CMS \cite{Chatrchyan:2012ufa} was a milestone in
modern particle physics and has marked 
a further great success of the
Standard Model (SM) describing the fundamental interactions of
elementary particles. However, there are still open questions that
cannot be answered within the SM such as the baryon-antibaryon
asymmetry. Sakharov proposed three 
conditions a model needs to fulfill in order to explain the dynamical
production of the baryon asymmetry through electroweak baryogenesis in the early
universe \cite{Sakharov:1967dj}. While the necessary baryon-violating processes are
possible in the SM, the required CP-violation and departure from thermal
equilibrium are not sufficient in the SM
\cite{Kajantie:1993ag,Fodor:1994sj,Jansen:1995yg,Kajantie:1995kf} to
explain the observed baryon asymmetry of the universe \cite{Bennett:2012zja}. 
Therefore, physics beyond the SM (BSM) is needed in
order to explain the baryon-anti-baryon asymmetry.\s

The requirement of departure from thermal equilibrium can be translated
into the requirement of a strong first order electroweak phase transition
(SFOEWPT). The presence of an SFOEWPT can be used as an additional
theoretical constraint a benchmark scenario of a BSM model needs to
fulfill. Only those parameter points that provide an SFOEWPT, are 
suitable candidates for electroweak baryogenesis. A minimalistic approach
to enable an SFOEWPT is the extension of the scalar sector of
the SM by including additional $SU(2)_L$ doublets or singlets in the
scalar potential. Despite the rich phenomenology of these models, there is an
increasing difficulty to simultaneously fulfill all experimental
constraints and the requirement of strong first order phase
transitions, see {\it e.g.}~\cite{Basler:2019iuu}. \s

On the other hand, there are more involved model frameworks such as
supersymmetry (SUSY) that solve (some of) the problems of the SM by a
symmetry relating bosonic and fermionic degrees of freedom. However,
theories like the Minimal Supersymmetric SM (MSSM) 
have come under strong experimental pressure in the past
years. In particular, LHC searches for colored scalars are
pushing the soft-SUSY-breaking scale of top superpartners above the
TeV-scale \cite{Sirunyan:2018vjp,Canepa:2019hph}. However, 
fermions charged under $SU(2)_L \otimes U(1)_Y$ are much less constrained and
currently still allowed to be as light as \unit[100-200]{GeV}
\cite{Aaboud:2017leg}.
This lead to a paradigm shift in the construction of SUSY-breaking
mechanisms and the proposal of \textit{split SUSY} scenarios
\cite{Wells:2003tf,Giudice:2004tc,ArkaniHamed:2004yi,Arvanitaki:2005fa,Bhattacharyya:2012ct}, 
which decouple all scalar superpartners of the SM fermions but maintain
rather light non-SM fermions at the same time. The resulting low-energy
effective field theory (EFT) often consists of a 
2-Higgs-Doublet-Model (2HDM) or a 2HDM with additional light
electroweak fermions and/or additional singlet fields.
\s

The (power-suppressed) contributions from heavy superpartners to
observables such as Higgs boson masses thereby can be incorporated
by performing a matching of a given effective field theory (EFT) Lagrangian on a more
fundamental model such as the MSSM 
\cite{Haber:1993an,Bagnaschi:2014rsa,Lee:2015uza,Bagnaschi:2017xid,Bahl:2018jom,Braathen:2018htl,Gabelmann:2018axh}. 
This strategy is currently also applied to Dark Matter observables 
\cite{Bhattacharyya:2012ct,Fox:2014moa,Gabelmann:2019jvz,Co:2021ion}, EWPTs
\cite{Athron:2019teq}, gravitational waves \cite{Demidov:2017lzf}
and electroweak baryogenesis \cite{Demidov:2006zz,Demidov:2016wcv}
in low energy EFTs of the Next-to-MSSM (NMSSM).
However, in contrast to Higgs boson mass predictions, the theoretical
accuracy in the matching for the other mentioned observables is often
only performed at tree-level and partially at one-loop
level. Furthermore, studies of the EWPT do 
not always consider an extended fermion sector but rather focus on the
extended scalar sector of the low-energy theory.
Neglecting the additional supersymmetric fermions can in
principle be justified in two different 
ways: (i) higgsinos/gauginos can be easily decoupled by assuming a large
$\mu$-term and large soft-breaking masses; (ii) the phenomenology of the EWPT is
determined by the shape of the scalar potential which is independent of the
fermion sector at leading order (LO). However, it was shown in Ref.
\cite{Gabelmann:2019jvz}, that the decoupling of fermions in non-minimal
spit-SUSY is not necessarily radiatively stable. Therefore, argument (i) cannot
always be applied. Furthermore, temperature corrections to the scalar
potential \textit{are} of higher order and therefore the
temperature profile of a scalar potential is also affected by light
fermions. As for (ii), in non-minimal Higgs sectors it might be
possible to compensate the effects of light fermions
by changing the many parameters in the scalar sector accordingly without
affecting any of the experimental constraints. However, this depends on
the considered model and therefore should be justified case-by-case.
\s

In this work we want to answer the question to which extent one of the simplest 
extensions of the SM which still features a strong first order EWPT and can
be embedded into split SUSY. We consider a 2HDM with additional
light $SU(2)_L \otimes U(1)_Y$ fermions and compute the strength of the EWPT
taking into account temperature corrections due to the additional
fermions. We compare the results with those obtained in the ordinary 2HDM and 
show that the fermion contributions can be large
enough to turn an EWPT that would be weak in the pure 2HDM, into a
strong first order EWPT. Furthermore, 
we demonstrate that in particular the parameter region of relatively heavy non-SM
Higgs bosons can benefit from the additional fermions.
\s

The paper is structured as follows. We introduce the model in
\cref{sec:2hdm}. The finite temperature corrections induced by the
additional fermion degrees of freedom are discussed in \cref{sec:tcorr}.
Numerical results of the strength of the EWPT for the model are shown in
\cref{sec:results} and compared to results obtained in an ordinary
2HDM without additional fermions. A short outlook on possible split
SUSY models is given \cref{sec:splitsusy}. We conclude in \cref{sec:conc}.

\section{A 2HDM with Electroweakinos}
\label{sec:2hdm}
In this paper, we work in the softly-broken $\mathbb{Z}_2$
symmetric version of the 2HDM \cite{Lee:1973iz,Branco:2011iw}
\begin{align}
    V_\text{2HDM}= &\,\, \frac{\lam1}{2} |\p1|^4 +  \frac{\lam2}{2} |\p2|^4 + \lam3 |\p1|^2|\p2|^2 + \lam4 |\p2^\dagger\p1|^2\nonumber\\
                   &\,\, +\left(\frac{\lam5}{2} \left(\p1^\dagger\p2\right)^2 -\ms{12}\p1^\dagger\p2 + h.c. \right) 
                     +\ms1|\p1|^2+\ms2|\p2|^2 \,,\label{eq:v2dhm} 
\end{align}
of type II, i.e. $\p1/\p2$ separately couple to the down/up-type
sector of the SM fermions. All parameters in the Higgs potential in
\cref{eq:v2dhm} are real and hence there is no explicit CP-violation
present in the Higgs potential. It is useful to define a basis
with two Higgs doublets $\Hu$ and $\Hd$ with different isospin, 
\begin{equation}
    \Hu = \p2 , \qquad \Hd = - i \sigma_2 \p1^*\,,
\end{equation}
such that $\Hu$, $\Hd$ and $\tan\beta={\langle \Hu \rangle}/ {\langle
\Hd \rangle}$ coincide
with the corresponding tree-level quantities of the MSSM. We have
introduced here the mixing angle $\tan \beta$ as the ratio of the two
vacuum expectation values of the neutral CP-even components of the
Higgs doublets, denoted by the square brackets. \s

We extend the default particle content of the real CP-conserving 2HDM by
additional non-SM like fermions. The BSM fermion sector should resemble the additional
non-SM fermion part of the MSSM. Therefore, we will refer to the
new fermions as electroweakinos in the following.
The Lagrangian for the BSM fermions reads
\begin{align}
    -\LagBSM = 
    & \,\, \frac{1}{\sqrt{2}}\Hu^\dagger\left(g_{2u} \sigma_a \tW^a + g_{1u}\tilde{B}\right)\tHu 
     \,\,-\frac{1}{\sqrt{2}}\Hd^\dagger\left(g_{2d} \sigma_a \tW^a + g_{1d} \tilde{B}\right)\tHd \nonumber \\ 
    & + \frac{M_{\tilde{W}}}{2} \tW^a \tW^a + \frac{M_{\tilde{B}}}{2} \tB\tB + \mu \tHu (i \sigma_2) \tHd 
      + h.c.\,\, , 
      \label{eq:lagfermion} 
\end{align}
where the bino $\tilde{B}$ is gauged under $U(1)_Y$ and the winos (higgsinos) 
$\tW$ ($\tHu,\tHd$) are triples (doublets) under $SU(2)_L$. 
The Pauli-matrices are referred to as $\sigma_a~(a=1,2,3)$. 
In contrast to the MSSM, the Yukawa couplings $g_{ij}$ with $i=1,2$
and $j=d,u$, and the Majorana masses $M_{\tilde{W}},\, M_{\tilde{B}},\, \mu$ are free
input parameters. Note however, for 
$g_{1u}=g_{1d}=g_1^\text{MSSM}$ and $g_{2u}=g_{2d}=g_2^\text{MSSM}$,
\cref{eq:lagfermion} coincides with the corresponding part of the tree-level
Lagrangian of the MSSM.
\s

Diagonalizing \cref{eq:lagfermion} yields four neutral mass eigenstates
$\tilde\chi_i^0~(i=1,...,4)$ with masses $m_{\tilde\chi^0_1}\le \dots
\le m_{\tilde\chi^0_4}$ and two 
charged mass eigenstates $\tilde\chi_j^-~(j=1,2)$ with masses $m_{\tilde\chi^-_1}\le
m_{\tilde\chi^-_2}$,
\begin{equation}
    \label{eq:inomass}
    \begin{aligned}
        \textbf{m}_{\tilde\chi^0} & = N^* \left(
\begin{array}{cccc}
    M_{\tilde{B}} & 0   					& -\frac{\vev\, g_{1d}}{2}\cb & \frac{\vev\, g_{1u}}{2}\sb \\[0.2cm]
        & M_{\tilde{W}} 					& \frac{\vev\, g_{2d}}{2}\cb  & -\frac{\vev\, g_{2u}}{2}\sb \\[0.2cm]
        &                       & 0 	                      & -\mu\\[0.2cm]
        &  \multicolumn{1}{c}{\text{\smash{\raisebox{1.5ex}{sym.}}}}                     &                             & 0 
\end{array}
\right) N^\dagger 
    = \text{diag}(m_{\tilde\chi_1^0},...,m_{\tilde\chi_4^0}) \\[0.4cm]
\textbf{m}_{\tilde\chi^{-}} & = U^* \left(
\begin{array}{cc}
    M_{\tilde{W}}                                & \frac{\vev\, g_{2d}}{\sqrt{2}} \sb \\[0.2cm]
    \frac{\vev\, g_{2u}}{\sqrt{2}} \cb & \mu
\end{array}
\right) V^\dagger
    = \text{diag}(m_{\tilde\chi_1^-},m_{\tilde\chi_2^-}) \,\, ,
\end{aligned}
\end{equation}
where $N$, $U$ and $V$ are unitary matrices and $\cb\equiv\cos\beta,
\sb\equiv\sin\beta$. The abbreviation 'sym.' indicates that the matrix
is symmetric with the corresponding entries to be filled in.

\section{Finite Temperature Corrections}
\label{sec:tcorr}
For the determination of the strength of the electroweak phase
transition, we use the loop-corrected effective potential at finite
temperature and follow the approach of
\cite{Basler:2016obg,Basler:2017uxn,Basler:2019iuu}. 
The loop-corrected effective potential $V(T)$ at finite temperature splits into
\begin{equation}
    V(T) = V_\text{tree} + \VCW + \VT + V_{\text{CT}}\,,
	\label{eq:v}
\end{equation}
where $V_\text{tree}$ is the tree-level potential and given by
\cref{eq:v2dhm}. The Coleman-Weinberg potential $\VCW$ is
the one-loop effective potential at zero temperature
\cite{Weinberg:1973am,Coleman:1973jx} and $\VT$
incorporates the temperature-dependent corrections
\cite{Dolan:1973qd,Quiros:1999jp}. The counterterm
potential $V_{\text{CT}}$ is defined such that the next-to-leading
order scalar (NLO) masses of the  
Higgs sector and the respective mixing angles are equal
  to the respective tree-level
input parameters used in the parameter scan. This allows for an
efficient parameter scan. A detailed discussion of the applied
renormalization scheme is given in \cite{Basler:2016obg}. Furthermore,
the derivation of the Coleman-Weinberg and of the temperature-dependent
potential of the scalar sector is equivalent to the pure 2HDM and
is presented in Ref.~\cite{Basler:2016obg}.
Therefore, here we only discuss those parts of
the calculation that change with respect to the pure 2HDM. 
The additional BSM fermions are taken into account by extending the
summation of all degrees of freedom in the Coleman-Weinberg potential
$\VCW$ and the temperature dependent potential $\VT$, respectively. The
required zero-temperature masses $m_{\tilde{\chi}^{(0,\pm)}_i}$ are obtained by
numerically diagonalizing the mass matrices in \cref{eq:inomass}. The
dominant thermal corrections, the Daisy corrections, are treated in the
Arnold-Espinosa method \cite{Arnold:1992rz,PhysRevD.46.2628}.
These corrections are taken into account by replacing the temperature
dependent potential with 
\begin{align}
    V_{\text{T}} \rightarrow V_{\text{T}} + \VD\,,
\end{align} 
with the daisy or Debye corrections
\begin{align}
    \VD =- \frac{T}{12\pi} \left[\sum\limits_{i=1}^{n_\mathrm{Higgs}} \left( \left( \overline{m}_i^2\right)^{3/2} - \left( m_i^2\right)^{3/2} \right)
    + \sum\limits_{a=1}^{n_\mathrm{gauge}} \left( \left( \overline{m}_a^2\right)^{3/2} - \left( m_a^2 \right)^{3/2} \right) \right] \,,
\end{align}
where the sum over the gauge bosons only extends over their
  longitudinal degrees of freedom.
The tree-level masses $m_{i/a}^2$ are the eigenvalues obtained from the
mass matrices $\Lambda_{(S)}^{ij}$ and $\Lambda_{(G)}^{ab}$ of the
scalars $S$ and SM gauge bosons $G$, respectively\footnote{We 
refer to Ref.~\cite{Camargo-Molina:2016moz} for a detailed description of the
notation of the mass matrices. }. The thermal masses
$\overline{m}^2_{i/a}$ are obtained by diagonalizing the mass matrices
while including the respective thermal ring corrections $\Pi_{(S)}$ and
$\Pi_{(G)}$\footnote{The explicit formulae for $\Pi_{(S)}$ and
$\Pi_{(G)}$ can be found in Ref.~\cite{Basler:2018cwe}.},
\begin{subequations}
	\begin{align} 
		\Lambda_{(S)}^{ij} &\to \Lambda_{(S)}^{ij} + \Pi_{(S)}^{ij} \,,\\
		\Lambda_{(G)}^{ab} &\to \Lambda_{(G)}^{ab} + \Pi_{(G)}^{ab} \,.
	\end{align}
\end{subequations}

\noindent The additional fermion contributions to $\Pi_S$ and $\Pi_G$
are given by
\begin{equation}
\Pi_{\Phi_i\to\Phi_j}^{\text{inos}} = \frac{T^2}{6}\left[
                                           \delta_{i1}\delta_{j1}\left(g_{1d}^2+3g_{2d}^2\right) 
                                         + \delta_{i2}\delta_{j2}\left(g_{1u}^2+3g_{2u}^2\right)
                                     \right],\quad\qquad 
\Pi_{W_i\to W_j}^{\text{inos}} = \delta_{ij} g_2^2 \frac{T^2}{2}, \quad\qquad 
\Pi_{B\to B}^{\text{inos}} = g_1^2 \frac{T^2}{6}, \label{eq:daisy}
\end{equation}
where $g_1$ and $g_2$ are the $U(1)_Y$ and $SU(2)_L$ gauge couplings.
The temperature-dependent corrections due to the fermions read
  \begin{align}
  V_{\text{T}}^{\text{inos}} = & 
          -\frac{T^4}{\pi^2} \text{Tr} \left[ 
               J_+\left(\textbf{m}^2_{\tilde\chi_i^0}/T^2\right) 
            +2 J_+\left(\textbf{m}^2_{\tilde\chi_i^-}/T^2\right)
  		\right] + \VD^{\text{inos}} 
  \label{eq:vt}
  \end{align} 
  with the thermal integral
  \begin{equation}
      J_\pm(x) = \int_0^\infty dk k^2 \log \left[1\pm
      \exp\left(-\sqrt{k^2+x} \right) \right]
      \, ,
      \label{eq:tint}
  \end{equation}
where the '-' ('+') sign refers to bosons (fermions).
In the decoupling limit of the electroweakinos 
\begin{equation}
M_{\tilde{B}},M_{\tilde{W}},\mu\to\infty\quad \text{and}\quad g_{ij}\to 0,
\label{eq:deoup}
\end{equation}
one can deduce from \cref{eq:vt,eq:tint,eq:daisy} that all temperature corrections
are heavily suppressed except for the ring corrections to the
longitudinal degrees of freedom 
of the $SU(2)_L$ and $U(1)_Y$ gauge bosons. However, the derivation of the ring
corrections \cite{PhysRevD.45.2933} 
assumes the high-temperature limit $T\gg m_i$. In fact, the ring corrections vanish 
by construction in the decoupling regime. Therefore, the limit \eqref{eq:deoup} has 
to be taken with care when performing a numerical study.
Another critical limit is the one of large Yukawa couplings $g_{ij}\gg 1$.
In this regime the Daisy resummation might also not be reliable anymore.
Therefore, we expect the most reliable results for intermediate
fermion masses and small and intermediate Yukawa couplings.
\s

We utilize the computer code \BSMPT
\cite{Basler:2018cwe,Basler:2020nrq} for the numerical study which was
extended to be able to calculate and minimize \cref{eq:v} in the presence of
additional non-SM like fermions. At the time of writing this publication, the extension
of \BSMPT is  private but we plan to release this feature in an upcoming
update of \BSMPT.

\section{Phenomenology of the EWPT}
\label{sec:results}
For the phenomenological analysis we start with the data sample
collected in Ref.~\cite{Basler:2016obg}, which studied the pure 2HDM
i.e.~\cref{eq:v2dhm} without the inclusion of \cref{eq:lagfermion}. 
The sample has been reevaluated using \ScannerS~{\tt 2.0.0} \cite{Coimbra:2013qq,Muhlleitner:2020wwk}
in order to apply the most recent collider constraints from $\higgsb$ \cite{Bechtle:2008jh,Bechtle:2011sb,Bechtle:2013gu,Bechtle:2013wla,Bechtle:2020pkv}
and $\higgss$\cite{Bechtle:2020uwn,Bechtle:2014ewa,Stal:2013hwa,Bechtle:2013xfa}. Using the obtained dataset, we investigate the impact of the additional non-SM-like fermions on the phenomenology of
the EWPT. The strength of the EWPT is given by the ratio of the electroweak VEV $\omega_c$ at the critical temperature $T_c$, 
\begin{equation}
    \xi_c\equiv \frac{\omega_c}{T_c}>1\,,
\end{equation}
which is required to be larger than one for a sufficiently strong first order EWPT. \s

In general, the presence of additional light fermions also affects
the Higgs boson branching ratios as well as other experimental
constraints such as the dark matter relic density. However, 
in this paper our focus is on the phenomenological impact of the fermions on the EWPT
rather than on collider physics and/or Dark Matter. A detailed study of the interplay
between the three aspects is reserved for future work.

\subsection{The Case $\xithdm<1$}
We start the discussion with a sample point that does not feature a
strong first order EWPT in the pure 2HDM, i.e. it has $\xithdm<1$ in the
absence of $\LagBSM$.
The input parameters  for the Higgs boson sector are
\begin{align}
        m_h &=\unit[125.09]{GeV},& m_H       &=\unit[637.37]{GeV}, \nonumber \\
        m_A &=\unit[811.35]{GeV},& m_{H^\pm} &=\unit[839.90]{GeV}, \label{eq:xilt1}\\
  \tan\beta &=6.15 \,\,,               &    \alpha &= -0.1605\,\, .\nonumber
\end{align}
The spectrum contains three rather heavy Higgs bosons above
\unit[600]{GeV} and the heavier neutral CP-even Higgs boson $H$ mixes only
by $\sim 2\%$ with the SM-like Higgs boson $h$.
Using these inputs we find for the strength of the EWPT
\begin{equation}
    \xithdm = 0.82\,,
\end{equation}
when considering the pure 2HDM type II.
\s

\begin{figure}[h]
    \centering
    \includegraphics[width=0.5\textwidth]{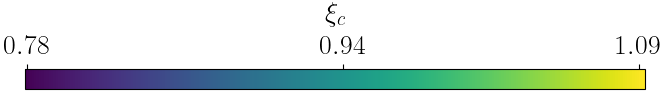}\\
    \includegraphics[width=0.49\textwidth]{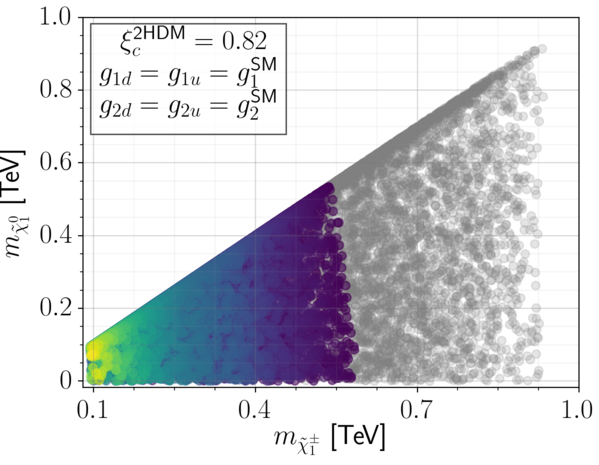}
    \includegraphics[width=0.49\textwidth]{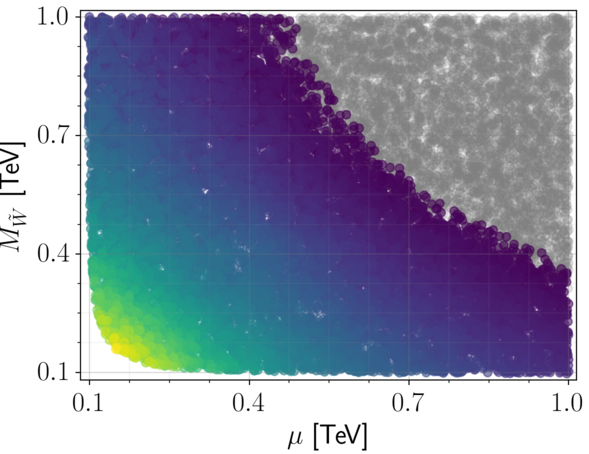}
    \caption{Prediction for $\xi_c$ for the parameter point defined in
    \cref{eq:xilt1} if only  $M_{\tilde{B}},M_{\tilde{W}}$ and $\mu$ are randomly varied.
Left: as a function of the two lightest electroweakino masses. Right: as
a function of $M_{\tilde{W}}$ and $\mu$.}
    \label{fig:massesx082}
\end{figure}

In the next step we assume $\LagBSM$ to be present and calculate the
strength of the EWPT for various scenarios of $g_{ij}$ and soft SUSY
breaking mass parameters $M_{\tilde{W}}$ $M_{\tilde{B}}$, and $\mu$.
The first scenario considers fixed Yukawa couplings 
that are given by the $U(1)_Y$ and $SU(2)_L$ SM gauge couplings, respectively,
$g_{1u}=g_{1d}=g_1^\text{SM}$ and $g_{2u}=g_{2d}=g_2^\text{SM}$ while
the soft SUSY breaking mass parameters are varied independently
between \unit[0.1-1]{TeV}. 
In \cref{fig:massesx082} we plot the determined $\xi_c$ values as a function
of the lightest charged and neutral electroweakino masses (left) and
as a function of the input parameters $M_{\tilde{W}}$ and $\mu$ (right). The
color code indicates the value of 
$\xi_c$ while grey points do not feature an NLO stable vacuum. We refer
to a scenario to have an NLO stable vacuum if the global
minimum of the one-loop effective potential at zero temperature
coincides with the normal electroweak vacuum present at tree level. 
In this and all further plots we take the model-independent constraint
$m_{\tilde{\chi}_1^-}>\unit[94]{GeV}$ from LEP data into account \cite{Abreu:1997ie}. 
\s

Figure~\ref{fig:massesx082} (left) indicates that additional light fermion
degrees of freedom are able to induce a strong first order EWPT with
$\xi_c>1$ for the considered parameter point which would otherwise
have a smooth phase transition in a pure 2HDM. Towards
large fermion masses, however, $\xi_c$ becomes smaller and for masses
$m_{\tilde\chi_1^{0,\pm}}\gtrsim \unit[600]{GeV}$ this benchmark scenario does not
provide an NLO stable vacuum any more. We point out that this sharp cut in
the mass plane is specific for this benchmark point and it is not
observed in all parameter points.\s

In \cref{fig:massesx082} (right) we see that also the $SU(2)_L$-specific
mass terms $M_{\tilde{W}}$ and $\mu$ show a strong correlation with $\xi_c$. On
the other hand, $M_{\tilde{B}}$ (not shown here) does not show a strong
correlation with $\xi_c$ in all investigated benchmark scenarios. \s

\begin{figure}[h]
    \centering
    {\hfill\includegraphics[width=0.435\textwidth]{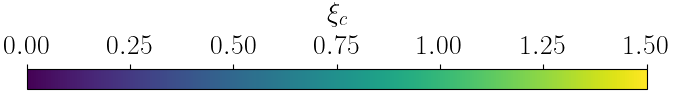}}\\
    \includegraphics[width=0.48\textwidth]{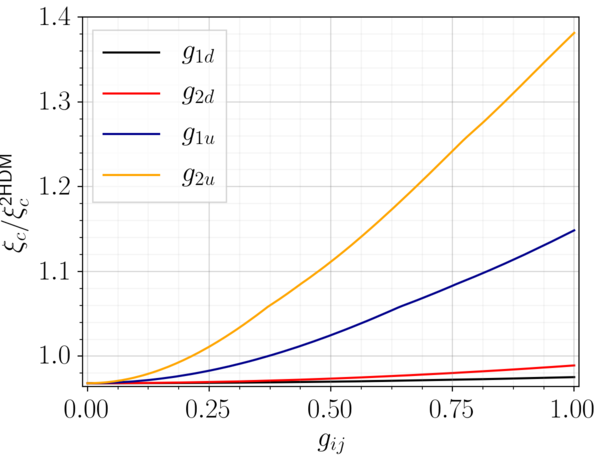}
    \includegraphics[width=0.5\textwidth]{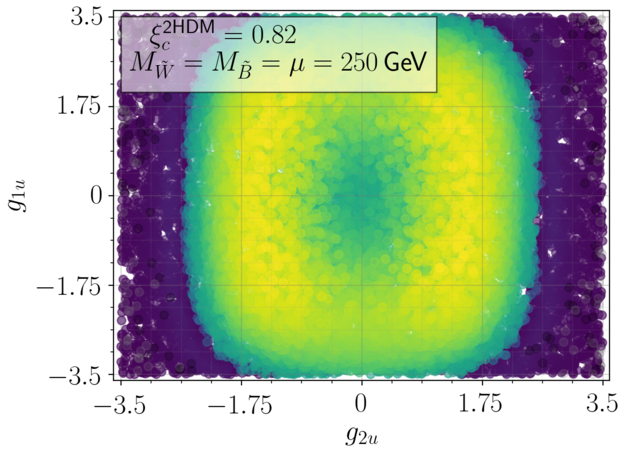}
     \caption{Prediction for $\xi_c$ for the parameter point defined in
    \cref{eq:xilt1} if only  $g_{1d},g_{2d},g_{1u}$ and $g_{2u}$ are varied.
    Left: all four couplings are varied individually. Right: all couplings
    are varied simultaneously.}
    \label{fig:yukawasx082}
\end{figure}

In the second scenario for $\xithdm<1$, we consider fixed soft
  SUSY breaking mass
parameters $M_{\tilde{W}}=M_{\tilde{B}}=\mu=\unit[250]{GeV}$ and study the impact of the
Yukawa couplings $g_{ij}$ on the EWPT. The variation of the Yukawa
couplings alters the masses of the new light fermions 
and hence also the loop-corrected effective potential as well as the thermal integrals in
\cref{eq:vt}. Furthermore, a change of the Yukawa couplings also
introduces a change in the 
additional temperature dependence through the Debye corrections
given in \cref{eq:daisy}.
We investigate the dependence of $\xi_c$ normalized to $\xithdm=0.82$
on the Yukawa couplings in \cref{fig:yukawasx082} when varying them
individually (left) from 0 to 1 and when varying them simultaneously
(right) between -3.5 and 3.5. 
The left plot in \cref{fig:yukawasx082} indicates that the Yukawa
couplings $g_{2i}$ ($i=u,d$) of the $SU(2)_L$ fermions (yellow and red
curves) have a stronger
impact on the strength of the phase transition $\xi_c$ than their
  corresponding $U(1)_Y$ 
couplings $g_{1i}$ (blue and dark green curves).
Overall the couplings to the up-type (i.e.~SM-like) Higgs boson have
a larger impact than those of the down-type Higgs boson (i.e. $\Phi_1$) with the 
largest impact of about 40\% for $g_{2u}=1$. The couplings to the
down-type Higgs boson have only a negligible effect.
The ratio $\xi_c/\xithdm$ does not tend to 1 in the limit
$g_{ij}\to0$ because there are still non-vanishing contributions to the
longitudinal vector boson, see \cref{eq:daisy}. \s

In the right plot of \cref{fig:yukawasx082} we vary all four Yukawa
couplings simultaneously. Similar to the $U(1)_Y$ mass parameter $M_{\tilde{B}}$, also
the down-type couplings $g_{1d}$, $g_{2d}$ (not shown here) have
only a minor impact on $\xi_c$. In contrast to the down-type couplings, 
the up-type couplings can shift $\xi_c$
from $\xi_c=0.8$ at $g_{iu}=0$ up to 1.4 for $g_{iu}<1$.
Allowing for even larger Yukawa couplings $g_{ij}>1$ could potentially spoil
perturbativity, however, and should be taken with care. \s

The interesting feature of the left plot
in \cref{fig:yukawasx082}, that the $SU(2)_L$ couplings have a
stronger impact than the corresponding  
$U(1)_Y$ couplings, is also seen in the right plot of
\cref{fig:yukawasx082}.
This behaviour can be explained with \cref{eq:daisy}, where the $SU(2)_L$
contributions have larger prefactors since they contribute with more
degrees of freedom. For very large Yukawa couplings above
$g_{2u}>2.5$, the strength of the EWPT $\xi_c$ is dropping rapidly to
0. \s

The capability of the additional fermions to strengthen the EWPT
compared to the 2HDM - or even pushing it into the strong regime - is not
unique to the considered parameter point. In \cref{fig:global}
(left), we show all points from the scan sample that are evaluated with the
extended fermion sector fixed by $g_{1u}=g_{1d}=g_1^\text{SM}$,
$g_{2u}=g_{2d}=g_2^\text{SM}$ and $M_{\tilde{B}}= M_{\tilde{W}}=\mu=\unit[200]{GeV}$.
The left plot shows the results in the ($m_H$,$m_A$)-plane when taking
the additional fermions into account. The grey points do not yield an NLO
stable electroweak vacuum, while the violet points develop an
electroweak vacuum with $\xi_c<1$. Green as well as yellow points
feature a strong first order EWPT, i.e. have $\xi_c>1$.
Yellow points are of particular interest since they do not
have a strong first order EWPT when considering the 2HDM without
extended fermion sector, i.e. $\xi_c>1$ but $\xithdm<1$.
The latter points favour regions with larger $m_A$ and $m_H$ values. In
this regime, the temperature-dependent corrections stemming from the
Higgs sector are too small to reach $\xithdm>1$ but the corrections from
the fermions are large enough to finally achieve $\xi_c>1$. 
\s

In summary, non-SM like fermion degrees of freedom can be a helpful
tool to weaken the tension between increasing Higgs boson mass
constraints from collider experiments and the requirement of additional
light scalar degrees of freedom to achieve a strong first order EWPT.

\subsection{The Case $\xithdm>1$}
\begin{figure}[t]
    \centering
    \includegraphics[width=0.49\textwidth]{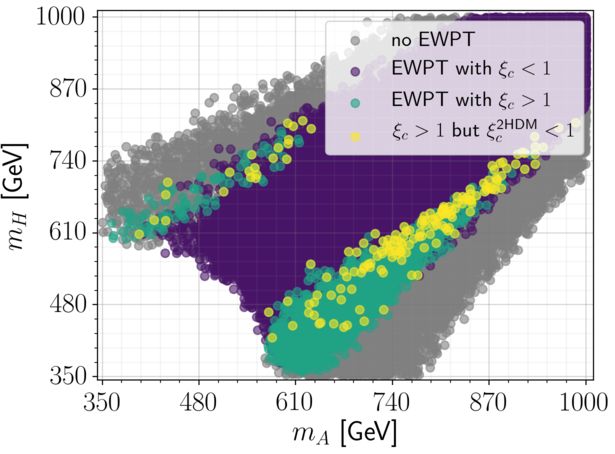}
    \includegraphics[width=0.49\textwidth]{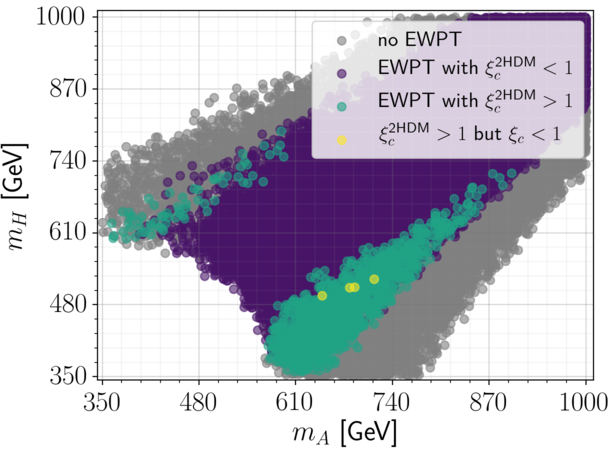}
    \caption{All points from the updated data sample of Ref. \cite{Basler:2016obg} in the $(m_H,m_A)$-plane.
    Left: evaluated with the extended fermion sector. Right: evaluated
    in the pure 2HDM.
    Grey points do not yield an NLO stable electroweak vacuum, while
    violet (green) points have a phase transi
tion with $\xi_c<1$
    ($\xi_c>1$), respectively. Yellow points
    in the left figure feature a strong first order EWPT in the model
    with an extended fermion sector but not in the pure 2HDM and vice versa
    for the right plot.
    }
    \label{fig:global}
\end{figure}

Another interesting case is if a strong first order
EWPT, $\xithdm>1$, is achieved in the 2HDM but not within the 2HDM with an
extended fermion sector. In the previous section, in
\cref{fig:massesx082,fig:yukawasx082} we have shown that 
additional fermions tend to strengthen rather then weaken the EWPT.
In \cref{fig:global} (right) we investigate if there can still be cases where $\xi_c<1$ with
additional fermions but $\xithdm>1$ in the pure 2HDM (yellow points). It shows the same
sample as in \cref{fig:global} (left) but evaluated in the 2HDM.
We find only few points for intermediate $(m_A,m_H)$ masses where
this is the case. In addition, these cases have
$\text{max}\{|\xi_c-\xithdm|\}\approx 10^{-3}$. This shows that the
overall tendency of weakening the EWPT through the BSM fermions is rather
tiny and hence negligible. 
In contrast, the yellow points in the left plot have
$\text{max}\{|\xi_c-\xithdm|\}\approx 0.4$. 


\section{Prospects for Split-SUSY}
\label{sec:splitsusy}
As already mentioned, the Yukawa interactions, $\LagBSM$, discussed in
\cref{sec:2hdm} coincide with the corresponding tree-level
Lagrangian of the MSSM. In addition, the considered type-II 2HDM can also
be matched to the 2HDM. However, in the MSSM we have $\lam5=0$ at
tree level. At the one-loop level, a matching of the MSSM to the 2HDM
including light electroweakinos generates additional operators
\begin{align}
    \left. V_\text{2HDM}\right|_\text{loop}  &= 
       \p1^\dagger\p2 \left(
        \frac{\lam5}{2}\p1^\dagger\p2 
        + \lam6\p1^\dagger \p1
        + \lam7\p2^\dagger\p2
        \right) + h.c.\,\,.  
    \end{align}
Integrating out heavy stops from the MSSM yields $\lam{5,6,7} \propto
(4\pi)^{-2}A_t^2/\msusy^2$, where $A_t$ is the trilinear
soft SUSY-breaking top Yukawa coupling. In split-SUSY, the soft-breaking stop
masses $\msusy$ are above the TeV
scale and are currently 
bounded from below to roughly \unit[1-2]{TeV}
\cite{Sirunyan:2018vjp}. On the other hand, $A_t$ 
is at the electroweak scale since it is protected by the 
same symmetry as the soft SUSY-breaking gaugino masses
\cite{Giudice:2004tc}.
Therefore, the loop-induced couplings cannot be much larger than
$\lam{5,6,7} \approx 10^{-3}$. 
However, we find in our 2HDM scan $\lam5\ge 0.1$ to
be a necessary condition for a strong first order EWPT if all other other quartic
couplings smaller than one (i.e. of similar order as in the MSSM). Considering that
all other SUSY degrees of freedom, except for the electroweakinos, are too
heavy to take part in the EWPT,
one can exclude the possibility of a strong first order EWPT, since the
effect of light electroweakinos is too small to compensate such a large
discrepancy.
This is also in agreement with the findings of Ref. \cite{Carena:2008vj}
which excluded a strong first order EWPT in the MSSM for stop masses
above \unit[115]{GeV}.
\s

However, in non-minimal SUSY models such as the
$\mathbb{Z}_3$-breaking\footnote{In general one can also generate
a non-zero tree-level $\lam5$ in the scale-invariant NMSSM. However, this
necessarily leads to a suppressed effective $\mu$-term,
$\mu\to 0$ for $m_\text{singlet}\to\infty$ and is
therefore not compatible with mass constraints on higgsinos.}
NMSSM it is possible to generate sizeable quartic couplings
$\lam{5,6,7}$ already at tree level by integrating out additional heavy
singlets. This would enable to achieve a strong first order EWPT in SUSY for
large stop and singlet masses while maintaining an effective and economic
2HDM near the SM scale. Furthermore, if the additional singlet
fields are light, the low-energy theory might coincide with a
$\mathbb{Z}_2$ breaking version of the singlet extended 2HDM
\cite{Athron:2019teq} or a singlet extended SM
\cite{Demidov:2006zz,Gabelmann:2018axh}.
While it is well known that additional light scalar singlets enrich the
EWPT phenomenology, the effect of additional light fermions that
are necessarily introduced when these models are motivated by split
SUSY, has so far not been studied in great detail.
In this paper, we have shown that the effect of light fermions on the
strength of the EWPT is in general not negligible. However, it is still
an open question whether the contributions from the fermions can be
compensated by changes in other regions of the relatively large
parameter space of non-minimal SUSY models. Therefore, a natural
next step is to study the EWPT generated by non-minimal extended Higgs
sectors in conjunction with extended fermion sectors, which we reserve
for future works.

\section{Conclusions}
\label{sec:conc}
We studied the impact of additional light $SU(2)_L$ and $U(1)_Y$ fermions on
the strength $\xi_c=v_c/T_c$ of the electroweak phase
transition in the type-II 2HDM 
using an extended version of the computer code \BSMPT.
It was shown that $\xi_c$ can significantly differ from the value
$\xithdm$ of the corresponding 2HDM that does not involve additional
fermions. Overall the strength of the EWPT changes in the presence
of additional fermions by up to 40\% compared to an ordinary 2HDM. 
The largest contribution is due to Yukawa couplings that couple
$SU(2)_L$ fermions to the up-type Higgs boson while bino-like states do have
a large impact.
Therefore, additional light fermions are a useful tool to weaken the
tension between increasing mass constraints on BSM scalars and the
requirement of additional light scalar degrees of freedom to accommodate
a strong first order EWPT. 
\s

The findings can be of particular importance for non-minimal split SUSY
models. While a strong first order EWPT in the (split) MSSM is already excluded,
models beyond the MSSM can yield a 2HDM with additional light
fermions - but enhanced quartic couplings - as a low-energy EFT.
The impact of non-minimal extended Higgs \textit{and}
fermion sectors is an effect yet to be scrutinized.
\section*{Acknowledgements}
M.G., M.M. and J.M. acknowledge support by the BMBF-Project 05H18VKCC1,
project number 05H2018. 

\bibliography{lit}

\end{document}